\documentclass{aa}  

\usepackage{graphicx}
\usepackage{txfonts}
\usepackage[]{hyperref}
\usepackage{graphicx}	
\usepackage{amsmath}	
\usepackage{amssymb}	
\usepackage{multicol}        
\usepackage{bm}		
\usepackage{pdflscape}	
\usepackage{threeparttable}
\newcommand\ktwo{\emph{{\it K2}}}
\newcommand{\pname}{K2-141}
\newcommand{\ename}{EPIC\,246393474}
\newcommand{\ms}{$\mathrm{m\,s^{-1}}$}

\newcommand{\kms}{$\mathrm{km\,s^{-1}}$}

\newcommand\kepler{\emph{{\it Kepler}}}
\newcommand\vsini{$v$\,sin\,$i_\star$}

\newcommand\teff{$T_{\rm eff}$}

\newcommand\logg{log\,{\it g$_\star$}}

\newcommand{\smass}[1][$M_{\odot}$]{$0.662\,\pm\,0.022$ #1} 
\newcommand{\sradius}[1][$R_{\odot}$]{$0.674\,\pm\,0.039$ #1}
\newcommand{\stemp}[1][$\mathrm{K}$]{ $4373\,\pm\,57$ #1 }
\newcommand{\Tzerob}[1][days]{$7738.46540 _{ - 0.00021 } ^ { + 0.00023 } $#1} 
\newcommand{\Pb}[1][days]{$0.2803226 \pm 0.0000013 $ #1} 

\newcommand{\bb}[1][]{$-0.01 _{ - 0.35 } ^ { + 0.38 }$ #1}       
\newcommand{\arb}[1][]{$2.31 _{ - 0.19 } ^ { + 0.08 }$ #1}       
\newcommand{\rrb}[1][]{$0.02088 _{ - 0.00039 } ^ { + 0.00053  }$ #1}       
 
\newcommand{\ib}[1][deg]{$ 90 \pm 10 $ #1} 
\newcommand{\ab}[1][AU]{$ 0.00716 _{ - 0.00065  } ^ {+ 0.00055 }$ #1}   
\newcommand{\densb}[1][$\mathrm{g\,cm^{-3}}$]{$ 2.98 _{ - 0.67 } ^ { + 0.33 }$  #1}
 
\newcommand{\rpb}[1][$R_{\oplus}$]{$1.54^{+0.10}_{-0.09}$\,#1}   
\newcommand{\denpb}[1][$\mathrm{g\,cm^{-3}}$]{$8.00 _{ - 1.45 } ^ { + 1.83  }$ #1}
\newcommand{\Tequib}[1][K]{$ 2039 _{ - 48 } ^ {+ 87 }$  #1}    
\newcommand{\ttotb}[1][hours]{$ 0.94 \pm 0.02 $  #1}
\newcommand{\qone}[1][]{ $0.54_{ - 0.25}^{ + 0.29 } $ #1}   
\newcommand{\qtwo}[1][]{ $0.29_{ - 0.19}^{ + 0.32 } $ #1}   
\newcommand{\uone}[1][]{ $0.43_{ - 0.26}^{ + 0.27 } $ #1}   
\newcommand{\utwo}[1][]{ $0.31_{ - 0.43}^{ + 0.36 } $ #1}   
 
\newcommand{\kbmone}[1][$m \, s^{-1}$]{$6.74 \pm 0.56  $ #1}  
\newcommand{\mpbone}[1][$M_{\oplus}$]{$5.31 \pm 0.46 $ #1}

\newcommand{\kbmtwo}[1][$m \, s^{-1}$]{$6.71 \pm 0.63 $ #1} 
\newcommand{\mpbtwo}[1][$M_{\oplus}$]{$5.23 \pm 0.50 $ #1} 

\newcommand{\kbmthree}[1][$m \, s^{-1}$]{$6.48^{+0.73}_{-0.71}$ #1} 
\newcommand{\mpbthree}[1][$M_{\oplus}$]{$5.05 _{ - 0.55  } ^ {+ 0.57 }$ #1}

\begin{document}

   \title{K2-141\,b:}
   \subtitle{A 5-$M_\oplus$ super-Earth transiting a K7\,V star every 6.7 hours\thanks{Based on observations obtained with \emph{a}) the Nordic Optical Telescope (NOT), operated on the island of La Palma jointly by Denmark, Finland, Iceland, Norway, and Sweden, in the Spanish Observatorio del Roque de los Muchachos (ORM) of the Instituto de Astrof\'isica de Canarias (IAC); \emph{b}) the 3.6m ESO telescope at La Silla Observatory under program ID 099.C-0491. \emph{c}) The \kepler\ space telescope in its extended mission \ktwo. }}

   \author{O.~Barrag\'an\inst{\ref{Torino}}, D.~Gandolfi\inst{\ref{Torino}}, F.~Dai\inst{\ref{cambridge},\ref{ivy}}, J.~Livingston\inst{\ref{tokyo}}, C.~M.~Persson\inst{\ref{chalmers}}, T.~Hirano\inst{\ref{tokyo2}}, N.~Narita\inst{\ref{tokyo},\ref{tokyo3},\ref{osawa}},
Sz.~Csizmadia\inst{\ref{DLR}},
J.~N.~Winn\inst{\ref{ivy}},
D.~Nespral\inst{\ref{lalaguna},\ref{IAC}},
J.~Prieto-Arranz\inst{\ref{lalaguna},\ref{IAC}}, 
A.~M.~S.~Smith\inst{\ref{DLR}},
G.~Nowak\inst{\ref{lalaguna},\ref{IAC}},  
S. Albrecht\inst{\ref{SAC}}, G.~Antoniciello\inst{\ref{Torino}}, A.~Bo~Justesen\inst{\ref{SAC}}, J.~Cabrera\inst{\ref{DLR}}, W.~D.~Cochran\inst{\ref{austin}}, H.~Deeg.\inst{\ref{lalaguna},\ref{IAC}}, Ph.~Eigmuller\inst{\ref{DLR}}, M.~Endl\inst{\ref{austin}}, A.~Erikson\inst{\ref{DLR}}, M.~Fridlund\inst{\ref{chalmers},\ref{Leiden}},
A. Fukui\inst{\ref{okayama}},
S.~Grziwa\inst{\ref{Koln}}, E.~Guenther\inst{\ref{TLS}}, A.~P.~Hatzes\inst{\ref{TLS}}, D.~Hidalgo\inst{\ref{lalaguna},\ref{IAC}}, 
M.C. Johnson\inst{\ref{ohio}},
J.~Korth\inst{\ref{Koln}}, E.~Palle\inst{\ref{lalaguna},\ref{IAC}}, M.~Patzold\inst{\ref{Koln}}, H.~Rauer\inst{\ref{DLR},\ref{TUBerlin}}, 
Y. Tanaka\inst{\ref{tokyo}},
and V.~Van~Eylen\inst{\ref{Leiden}}
    }

\institute{
	Dipartimento di Fisica, Universit\`a degli Studi di Torino, via Pietro Giuria 1, I-10125, Torino, Italy\label{Torino} \\ \email{oscar.barraganvil@edu.unito.it}
 \and Department of Physics and Kavli Institute for Astrophysics and Space Research, Massachusetts Institute of Technology, Cambridge, MA, 02139, USA \label{cambridge}
 \and Department of Astrophysical Sciences, Princeton University, 4 Ivy Lane, Princeton, NJ, 08544, USA \label{ivy}
 \and Department of Astronomy, Graduate School of Science, The University of Tokyo, Hongo 7-3-1, Bunkyo-ku, Tokyo, 113-0033, Japan \label{tokyo}
 \and Department of Space, Earth and Environment, Chalmers University of Technology, Onsala Space Observatory, 439 92 Onsala, Sweden \label{chalmers}
 \and Department of Earth and Planetary Sciences, Tokyo Institute of Technology, 2-12-1 Ookayama, Meguro-ku, Tokyo 152-8551, Japan \label{tokyo2}
 \and Astrobiology Center, NINS, 2-21-1 Osawa, Mitaka, Tokyo 181-8588, Japan \label{tokyo3}
 \and National Astronomical Observatory of Japan, NINS, 2-21-1 Osawa, Mitaka, Tokyo 181-8588, Japan \label{osawa}
 \and Institute of Planetary Research, German Aerospace Center, Rutherfordstrasse 2, 12489 Berlin, Germany\label{DLR}
\and Departamento de Astrofísica, Universidad de La Laguna, E-38206, Tenerife, Spain \label{lalaguna} 
\and Instituto de Astrofísica de Canarias, C/ Vía Láctea s/n, E-38205, La Laguna, Tenerife, Spain\label{IAC} %
\and Stellar Astrophysics Centre, Deparment of Physics and Astronomy, Aarhus University, Ny Munkegrade 120, DK-8000 Aarhus C, Denmark\label{SAC}
\and Department of Astronomy and McDonald Observatory, University of Texas at Austin, 2515 Speedway, Stop C1400, Austin, TX 78712, USA \label{austin}
\and Leiden Observatory, University of Leiden, PO Box 9513, 2300 RA, Leiden, The Netherlands\label{Leiden}
\and Okayama Astrophysical Observatory, National Astronomical Observatory of Japan, Asakuchi, Okayama 719-0232, Japan\label{okayama}
\and Rheinisches Institut f\"ur Umweltforschung, Abteilung Planetenforschung an der Universit\"at zu K\"oln, Aachener Strasse 209, 50931 K\"oln, Germany\label{Koln}
\and Th\"uringer Landessternwarte Tautenburg, Sternwarte 5, D-07778 Tautenberg, Germany\label{TLS}
\and Department of Astronomy, The Ohio State University, 140 West 18th Ave., Columbus, OH 43210, USA\label{ohio},
\and Center for Astronomy and Astrophysics, TU Berlin, Hardenbergstr. 36, 10623 Berlin, Germany\label{TUBerlin}
}
   
\date{Received October 31, 2017; accepted November 4, 2017}

\titlerunning{The ultra-short-period planet \ename\,b}
\authorrunning{Barrag\'an et al.}

  \abstract
   {                  
We report on the discovery of \pname\,b  (\ename\,b), an ultra-short-period super-Earth on a 6.7-hour orbit transiting an active K7\,V star based on data from \ktwo\ campaign 12.  We confirmed the planet's existence and measured its mass with a series of follow-up observations: seeing-limited MuSCAT imaging, NESSI high-resolution speckle observations, and FIES and HARPS high-precision radial-velocity monitoring.  \pname\,b has a mass of \mpbone\ and radius of \rpb, yielding a mean density of \denpb and suggesting a rocky-iron composition. Models indicate that iron cannot exceed $\sim$70\% of the total mass. With an orbital period of only 6.7 hours, \pname\,b is the shortest-period planet known to date with a precisely determined mass.
}


   \maketitle
%

\section{Introduction}

Short-period ($P_\mathrm{orb}\,\lesssim\,10$\,days) exoplanets are interesting and convenient targets for radial velocity (RV) follow-up observations. The shorter the period, the larger the amplitude of the Doppler reflex motion induced by the planet on its host star, and the easier it is to sample the orbital motion in an observing campaign. This helps to explain why the ultra-short-period (USP, P\,$< 1$\,days) planets have often been the targets of recent radial-velocity programs.  With periods shorter than one day, and sizes almost always smaller than $2\,R_\oplus$, transiting USP planets offer relatively easy access to knowledge of the properties of terrestrial-sized objects \citep{Sanchis2014}. The radius domain of USP planets includes the gap between 1.5\,--\,2\,$R_\oplus$ in the bimodal distribution found in the \kepler\ sample by \citet{Fulton2017}. \citet{2017arXiv171005398V} found a similar result using a small sample with stellar parameters coming from asteroseismology measurements, which includes transiting planets with radius determinations of $\sim3\%$. This gap has been predicted by photo-evaporation models \citep[e.g., ][]{Lopez2014,Owen2013}, in which close-in planets can lose their entire atmospheres due to stellar irradiation. In this context, USP planets are expected to be atmosphere-free, bare, solid planets. Accurate mass measurements of transiting USP planets will be helpful to test this theory. They may also help to determine whether there is substantial variation in the balance of rock and iron in terrestrial planets \citep[e.g.,][]{Pepe2013,Gandolfi2017,Guenther2017}.  

\begin{figure*}
\includegraphics[width=1\textwidth]{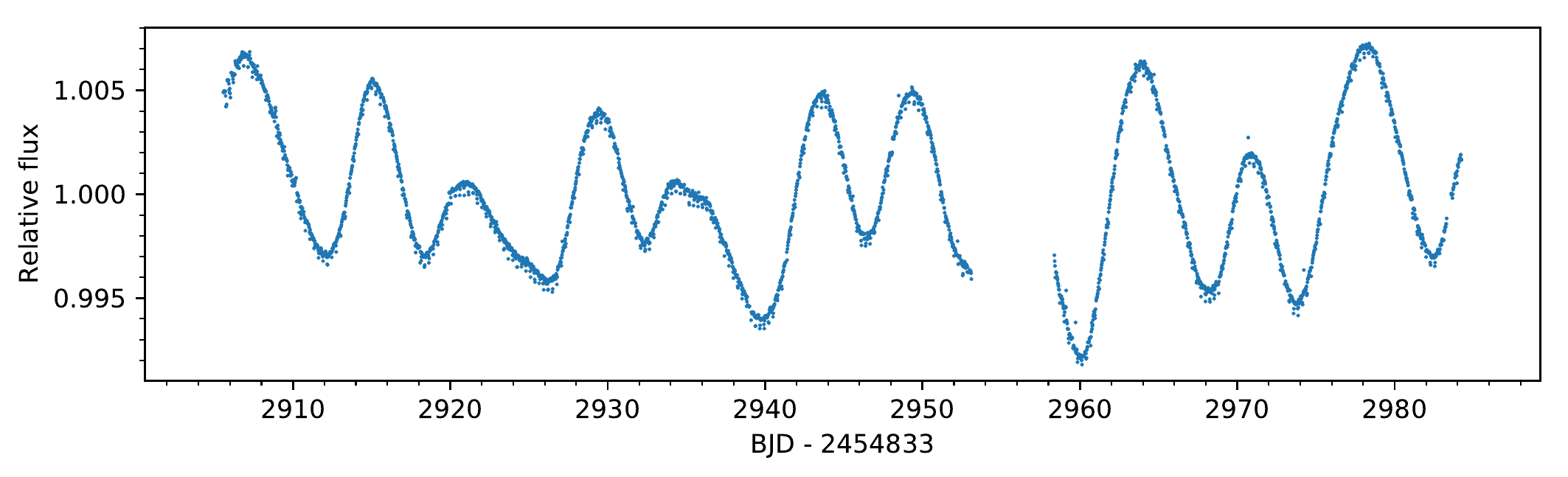}
\caption{\ktwo\ light curve of \pname. Stellar activity is seen as the quasi-periodic, long period modulation. Transits are visible as shallow dips. The 5.3-day-long data gap, during which the telescope entered in safe mode, is clearly visible at $\sim$2/3 of the time series.\label{Fig:K2LC}}
\end{figure*}

\citet{Sanchis2014} suggested that USP planets are the remnant cores of hot Jupiters that lost their gaseous envelopes due to photo-evaporation. However, \citet{Winn2017} found that the strong tendency for gas-giant planets to be found around metal-rich stars does not hold for USP planets, contrary to what one would expect if hot Jupiters are the progenitors of USP planets. This leaves open the possibility that hot Neptunes are the progenitors of USP planets, because the occurrence rate of planets smaller than $4\,R_\oplus$ does not appear to be strongly dependent on the host star metallicity \citep{Buchhave2012, Buchhave2015}. Accurate determinations of masses and radii of USP planets --- with uncertainties of 20\% or smaller --- along with spectral analyses of the host stars would enable searches for any possible correlations between stellar composition and the properties of USP planets \citep[e.g., ][]{Dumusque2014,Gandolfi2017}.
 
In this paper we present the discovery of \pname\,b (\ename\,b), an USP planet transiting a K7\,V star. The use of state-of-the-art spectrographs and an optimal observing strategy allowed us to pin down the planetary mass with an uncertainty better than 10\%. These results are part of the ongoing project carried out by the KESPRINT consortium to detect, confirm, and characterize transiting planets from the \ktwo\ mission  \citep[e.g., ][]{Barragan2016,Barragan2017,Dai2017,Fridlund2017,Johnson2016,Narita2017,Nespral2017,Sanchis-Ojeda2015,  Smith2017}.



\section{\ktwo\ photometry}
\label{Sect:K2Photometry}

\begin{table}
\caption{Main identifiers, coordinates, optical and infrared magnitudes, and proper motion of \ename.  \label{tab:parstellar} 
}
\begin{center}
\begin{tabular}{lcc} 
\hline
\hline
\noalign{\smallskip}
Parameter & Value &  Source \\
\noalign{\smallskip}
\hline
\noalign{\smallskip}
\multicolumn{3}{l}{\emph{Main identifiers}} \\
\noalign{\smallskip}
TYC  & 5244-714-1 & EPIC \\
EPIC & 246393474   & EPIC \\
UCAC & 445-136683 & EPIC \\
2MASS & 23233996-0111215 & EPIC \\
\noalign{\smallskip}
\hline
\noalign{\smallskip}
\multicolumn{3}{l}{\emph{Equatorial coordinates}} \\
\noalign{\smallskip}
$\alpha$(J2000.0) & $23^\mathrm{h}23^{\mathrm{m}}39.971^{\mathrm{s}}$ & EPIC \\
$\delta$(J2000.0) & -01$^{\circ}$ 11$^\prime$21.39${\arcsec}$ & EPIC \\
\noalign{\smallskip}
\hline
\noalign{\smallskip}
\multicolumn{3}{l}{\emph{Magnitude}} \\
$V_\mathrm{T}$ & 11.532$\pm$0.156 & Tycho \\
$V_\mathrm{J}$ & 11.389$\pm$0.148 & calc. \\
$I_\mathrm{C}$ & 9.967$\pm$0.030 & DENIS \\
$J$ &  9.086$\pm$0.021 & 2MASS \\
$H$ &  8.524$\pm$0.053 & 2MASS \\
$Ks$ &  8.401$\pm$0.023 & 2MASS \\
$W1$ & 8.311$\pm$0.024 & WISE \\
$W2$ & 8.391$\pm$0.020 & WISE \\
$W3$ & 8.311$\pm$0.023 & WISE \\
$W4$ & 7.930$\pm$0.223 & WISE \\
\noalign{\smallskip}
\hline
\noalign{\smallskip}
\multicolumn{3}{l}{\emph{Proper motions and parallax}} \\
$\mu_{\alpha} \cos \delta$ (mas \ yr$^{-1}$) & $38.584 \pm 3.907 $ & Gaia\\
$\mu_{\delta} $ (mas \ yr$^{-1}$) & $-9.837 \pm 3.534$ & Gaia \\
Parallax (mas) & $ 17.01\pm0.81$ & Gaia \\ 
\noalign{\smallskip}
\hline
\end{tabular}
\end{center}
\begin{tablenotes}\footnotesize
 \item \emph{Note} -- Values of fields marked with EPIC are taken from the Ecliptic Plane Input Catalog available at \url{http://archive. stsci.edu/k2/epic/search.php}. Values marked with Tycho, DENIS, 2MASS, WISE, and Gaia are from \citet{Hog2000}, \citet{Denis2005}, \citet{Cutri2003}, \citet{Cutri2013}, and \citet{Fabricius2016}, respectively. The Jonson V-band magnitude ($V_\mathrm{J}$) has been calculated from the Tycho $V_\mathrm{T}$ and 2MASS $H$ magnitudes using Eric Mamajek's transformation equation available at \url{https://doi.org/10.6084/m9.figshare.1291187.v1}.
\end{tablenotes}
\end{table}

\ktwo's campaign 12 (C12) was carried out by the \kepler\ space telescope from December 15, 2016 to March 4, 2017 UTC. The spacecraft was pointed towards the coordinates $\alpha_{\rm J2000}=23^{\rm h}26^{\rm m}38^{\rm s}$, $\delta_{\rm J2000}=-05^{\rm o} 06' 08''$. The photometric data include 79 days of almost continuous observations with a gap of 5.3 days between Feb 01, 15:06 UTC and Feb 06, 20:47 UTC when the satellite entered a safe mode. The C12 target list included 29,221 targets monitored in long-cadence mode, and 141 in short-cadence mode\footnote{See \url{http://keplerscience.arc.nasa.gov/k2-fields.html}.}. We downloaded the calibrated target pixel files from the Mikulski Archive for Space Telescopes.\footnote{https://archive.stsci.edu/k2.} and extracted the light curves using an approach similar to that described by \citet{Vanderburg2014}. For each image, we laid down a 16\arcsec--wide circular aperture around the brightest pixel, and fitted a 2-d Gaussian function to the intensity distribution. We then fitted a piecewise linear function between the observed flux variation and the best-fitting central coordinates of the Gaussian function. 

Before searching the light curves for transits, we attempted to remove long-term systematic or instrumental flux variations by normalizing the light curve using 1.5-day-long cubic splines. We searched for periodic transit signals using the Box-Least-Squares algorithm \citep[BLS,][]{Kovacs2002} and employed a non-linear frequency grid to account for the expected scaling of transit duration with orbital period. We also adopted \citet{Ofir2014}'s definition of signal detection efficiency (SDE). We discovered that the light curve of \ename\ (hereafter \pname) -- whose equatorial coordinates, proper motion, parallax, and optical and near-infrared magnitudes are listed in Table~\ref{tab:parstellar} -- shows periodic transit-like signal with a SDE of 23.5 and a depth of $\sim$0.04\% occurring every $0.28$\,days (6.7 hours). We searched for additional transiting planets in the system by re-running the BLS algorithm after removing the data within 1.5~hours of each transit of planet b. No transit signal was detected: the maximum SDE of the new BLS spectrum was 4.5. A visual inspection of the light curve did not reveal any additional transits, either. The target passed standard tests used to detect false positives due to eclipsing binaries: we did not detect any secondary eclipses or alternation of eclipse depths.

The {\it K2} light curve of \pname\ shows quasi-periodic variations with a peak-to-peak amplitude of about $\sim$1\%, very likely the result of rotation and active regions on the host star (Fig.~\ref{Fig:K2LC}). This will be further discussed in Sect.~\ref{Sect:Activity}. Using the auto-correlation method applied to the out-of-transit \ktwo\ light curve \citep{McQuillan2014}, we measured a stellar rotation period of $P_\mathrm{rot}=14.03\pm0.09$\,days. 

\section{Ground-based follow-up observations}

\kepler's CCDs have a sky-projected pixel size of $\sim$4$\arcsec$. Ground-based imaging with higher angular resolution is useful to check for an unresolved eclipsing binary that might be the source of the transit signal. Imaging is also useful to measure the fraction of light in the \ktwo\ photometric aperture that originates from the target star as opposed to other nearby stars. This is important for establishing the true fractional variation in the starlight during transits, and thereby the planet radius. Finally, high-precision RV measurements are needed to confirm the planetary nature of the transiting object and to measure its mass.

\subsection{Diffraction-limited imaging}
\label{Section:Speckle imaging}

\begin{figure}[!t]
\centering
\includegraphics[width=\linewidth]{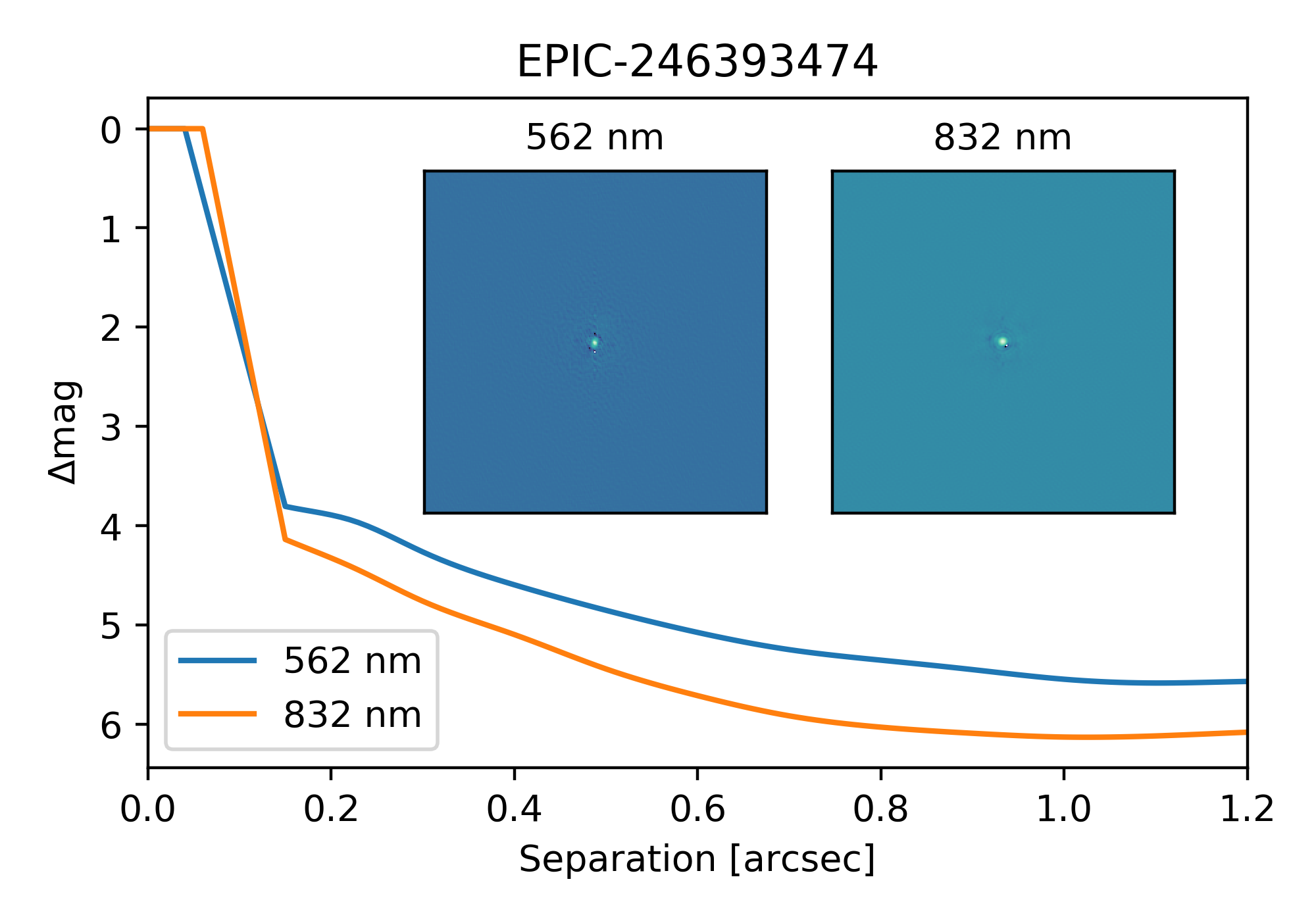}
\caption{``Blue'' and ``red'' contrast curves and reconstructed images of \pname (insets). The two images are both centered around \pname\ and have a size of 4.6$\arcsec$\,$\times$\,4.6$\arcsec$.} 
\label{Fig:C12_3474_ConstrastCurve}
\end{figure}

On the night of August 9, 2017 UT, we conducted speckle imaging observations of the star \pname\ with the NASA Exoplanet Star and Speckle Imager (NESSI; Scott et al., 2017, PASP, in prep.), a new instrument for the WIYN 3.5 meter telescope, which uses high-speed electron-multiplying CCDs (EMCCDs) to capture sequences of 40\,ms exposures simultaneously in two bands. We also observed nearby point source calibrator stars close in time to the science target. We observed simultaneously in the ``blue'' band centered at 562\,nm with a width of 44\,nm and the ``red'' band centered at 832\,nm with a width of 40\,nm. The pixel scales of the ``blue'' and ``red'' EMCCDs are 0.0175649 and 0.0181887 $\arcsec$/pixel, respectively. The data reduction followed the same procedures described by \citet{Howell2011}. Using the point source calibrator images, we computed reconstructed 256\,$\times$\,256 pixel images in each band, corresponding to ~4.6$\arcsec$\,$\times$\,4.6$\arcsec$. Figure~\ref{Fig:C12_3474_ConstrastCurve} shows the contrast curves and reconstructed images for the ``blue'' and ``red'' pass bands. No secondary sources were detected in the reconstructed images. We measured the background sensitivity of the reconstructed images using a series of concentric annuli centered on the target star, resulting in 5$\sigma$ sensitivity limits (in delta-magnitudes) as a function of angular separation.

\begin{table*}[!t]
\begin{center}
\caption{Spectroscopic parameters of \pname\ as derived from the HARPS data using the two methods described in Sect.~\ref{Sec:StellarParameters}.\label{Tab:SpecParam}}
\begin{tabular}{lccccccccc}
\hline
\hline
\noalign{\smallskip}
Method  &    \teff\ & \logg\  & [Fe/H] &  $R_\star$   &  \vsini\ \\
        & (K) &  (cgs)  &  (dex) &  ($R_\odot$) &  (\kms) \\
\noalign{\smallskip}
\hline
\noalign{\smallskip}
\texttt{SME V5.2.2}    & $4403\pm100$  & $4.50\pm0.15$ & $-0.06\pm0.20$ &   \dotfill    & $3.0\pm1.7$ \\
\texttt{SpecMatch-emp} & $4359\pm70$~~ & \dotfill      & $+0.06\pm0.12$ & $0.701\pm0.070$ &  \dotfill    \\
\noalign{\smallskip}
\hline
\end{tabular}
\end{center}
\end{table*}

\subsection{Seeing-limited imaging}
\label{Section:Muscat}

In order to search for sources outside of the 4.6$\arcsec$\,$\times$\,4.6$\arcsec$ field-of-view of our high-resolution NESSI images, but within the 16$\arcsec$--wide circular aperture used to extract the light curve from the K2 pixel files, we also obtained seeing-limited multi-band optical images using the MuSCAT \citep{Narita2015} instrument on the 1.88m telescope at the Okayama Astrophysical Observatory. The pixel scale of MuSCAT's CCDs is 0.36\,$\arcsec$/pixel. The instrument can observe in Sloan $g^\prime_2$, $r^\prime_2$, and $z_{s,2}$ bands simultaneously. The observations were performed on September 23, 2017 UT with seeing of $\sim$1.5\,$\arcsec$. To keep the peak count level at about 45,000 ADU, we took 10 exposures of 15, 4, and 12 sec with the $g^\prime_2$, $r^\prime_2$, and $z_{s,2}$ bands, respectively. The frames were dark-subtracted and flat-fielded using standard routines. The 5 best on-focus frames were stacked together for each band. 

We detected a companion located at 12.6 arcsec towards the East of \pname. MuSCAT's images show that this target is about 7.8 mag fainter than \pname\ (from the weighted averaged of the $g^\prime_2$ and $r^\prime_2$ filters). This accounts for a contamination factor of $1/1400$ of the target brightness. This value does not have a measurable impact on the derived parameters.

\subsection{High-precision Doppler observations}
\label{Section:spectroscopy}

We started the RV follow-up of \pname\ with the the FIbre-fed \'Echelle Spectrograph \citep[FIES;][]{Telting2014} mounted at the 2.56m Nordic Optical Telescope (NOT) of Roque de los Muchachos Observatory (La Palma, Spain). We obtained 9 high-resolution ($R\,\approx\,67\,000$) spectra on 5 different nights, from August 15 to September 14, 2017 UTC, within observing programs 55-019, 55-202, and 55-206. Since the 6.7-hour period is short enough to allow a significant fraction of the orbit to be sampled in one night, we obtained two spectra per night during 3 of the 5 FIES observing nights. Following \citep{Gandolfi2015}, we traced the RV drift of the instrument by bracketing the science exposures with 90-sec ThAr spectra. We reduced the data using standard IRAF and IDL routines and extracted the RVs via multi-order cross-correlations using the stellar spectrum with the highest S/N ratio as a template.

We also acquired 27 spectra with the HARPS spectrograph \citep[$R\,\approx\,115\,000$;][]{Mayor2003} mounted at the ESO-3.6m telescope of La Silla observatory (Chile), as part of the observing program 099.C-0491. We adopted the same observing strategy as the FIES observations, acquiring 2-5 spectra per night on 7 different nights, from August 19 to 27, 2017 UTC. We reduced the data using the dedicated off-line HARPS pipeline and extracted the RVs via cross-correlation with a K5 numerical mask. The pipeline provides also the bisector span (BIS) and full-width at half maximum (FWHM) of the cross-correlation function, along with the Mt. Wilson activity index $S$ of the Ca\,{\sc ii} H\,\&\,K lines. Table~\ref{RV} reports the FIES and HARPS RV measurements, as well as the exposure times and S/N ratios per pixel at 5500\,\AA.

\section{Stellar fundamental parameters}
\label{Sec:StellarParameters}

We derived the spectroscopic parameters of \pname\ from the co-added HARPS spectrum, which has a S/N ratio per pixel of $\sim$250. We used Spectroscopy Made Easy \citep[\texttt{SME};][]{vp96,vf05,Piskunov2017}, a spectral analysis package that calculates synthetic spectra and fits them to high-resolution observed spectra using a $\chi^2$ minimizing procedure. The analysis was performed with the non-LTE \texttt{SME} version 5.2.2, along with \texttt{ATLAS\,12} model atmospheres \citep{Kurucz2013}. The microturbulent and macroturbulent velocities were assumed to be 1\,\kms \citep{Gray2008}. The wings of the H$_\alpha$ and H$_\beta$ lines were used to measure the effective temperature \teff. We excluded the core of the Balmer line because of their origin in higher layers of stellar photospheres. The surface gravity \logg\ was determined from the wings of the Ca\,{\sc i}~$\lambda$\,6102, $\lambda$\,6122, $\lambda$\,6162\,\AA\ triplet, and the Ca\,{\sc i} $\lambda$\,6439\,\AA\ line. We measured the iron abundance [Fe/H] and projected rotational velocity \vsini\ by simultaneously fitting many unblended iron lines in the spectral region 5880--6600\,\AA.

An independent analysis was carried out with \texttt{SpecMatch-emp} \citep{2017ApJ...836...77Y}, a tool that uses hundreds of Keck/HIRES high-resolution template spectra of FGK stars for which the parameters have been accurately measured via interferometry, asteroseismology, spectral synthesis, and spectrophotometry. \texttt{SpecMatch-emp} finds the templates that best match an input observed spectrum in the spectral region 5000-5900\,\AA\ and derives the effective temperature $T_\mathrm{eff}$, stellar radius $R_\star$, and iron abundance [Fe/H] by interpolation.

\begin{figure}
\centering
\includegraphics[width=\linewidth]{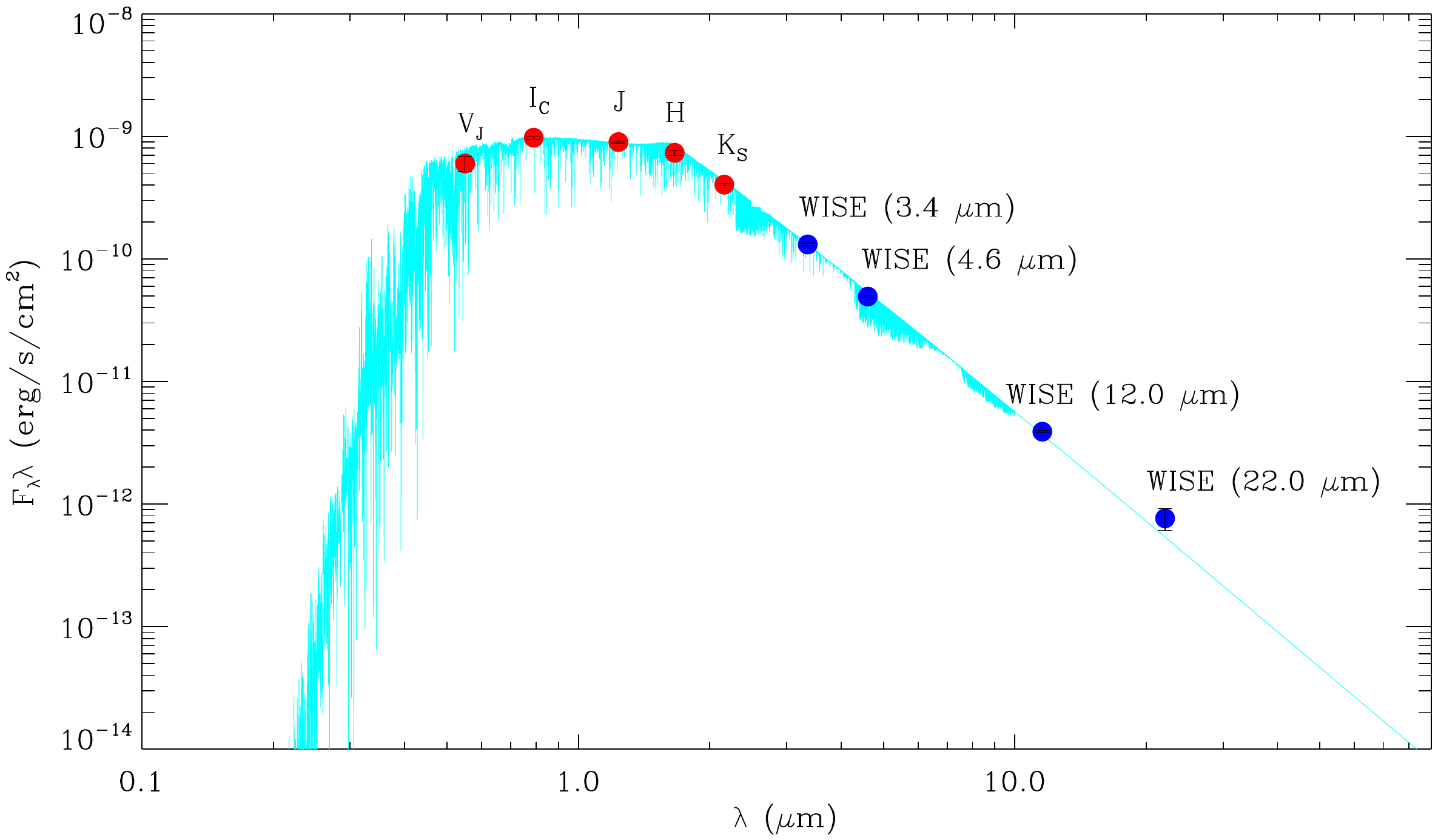}
\caption{Spectral energy distribution of \pname. The \texttt{BT-Settl-CIFIST} model spectrum with the same parameters as the star is plotted with a light blue line. The $V_\mathrm{J}$, $I_\mathrm{C}$, $J$, $H$, $Ks$, $W1$, $W2$, $W3$, and $W4$ fluxes are derived from the magnitudes reported in \autoref{tab:parstellar}.} 
\label{Fig:C12_3474_SED}
\end{figure}

\begin{figure*}[!th]
\centering
\includegraphics[width=\textwidth]{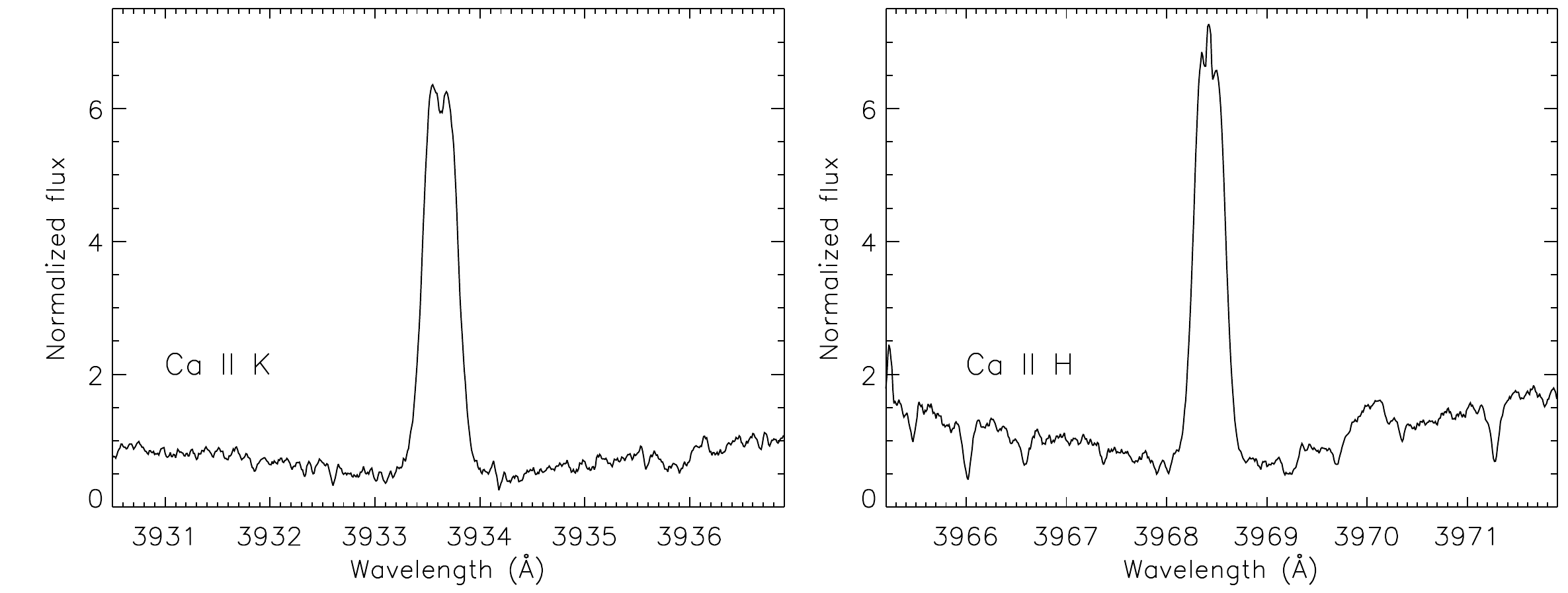}
\caption{Cores of the Ca\,{\sc ii} H\,\&\,K lines of \pname\ as observed with HARPS.} 
\label{Fig:CaHK}
\end{figure*}

We summarize the results of the two spectral analyses in Table~\ref{Tab:SpecParam}. The effective temperature and iron abundance estimates are consistent well within the nominal error bars. Since the two methods are based on different wavelength regions ($\lambda\,>\,5880$\,\AA\ for \texttt{SME} and $\lambda\,<\,5900$\,\AA\ for \texttt{SpecMatch-emp}) we treated the two sets of parameters as independent estimates. For \teff\ and [Fe/H], we computed the weighted means of the values derived from the two methods. For the projected rotational velocity \vsini\ we adopted the value determined with \texttt{SME}. We report the adopted effective temperature \teff, iron abundance [Fe/H], and projected velocity \vsini\ in Table~\ref{parstable1}. The stellar radius and surface gravity were re-determined using a different method, as described in the next paragraphs.

We derived the stellar radius $R_\star$ and reddening $A_\mathrm{v}$ following the method described in \citep{Gandolfi2008}. Briefly, we first built the spectral energy distribution (SED; Fig.~\ref{Fig:C12_3474_SED}) of \pname\ from the optical and infrared photometry listed in Table~\ref{tab:parstellar}. We then fitted the SED using the \texttt{BT-Settl-CIFIST} \citep{Baraffe2015} model spectrum with the same spectroscopic parameters as the star. Adopting the extinction law of \citet{Cardelli1989} and assuming a total-to-selective extinction of $R\,=\,A_\mathrm{v}/E_{B-V}\,=\,3.1$, we found that the interstellar reddening is consistent with zero ($A_\mathrm{v}$\,=\,0.01\,$\pm$\,0.02~mag). Using the distance of d\,=\,$58.77\,\pm\,0.81$\,pc from the {\it Gaia}'s first data release \citep[Table~\ref{tab:parstellar};][]{GaiaDR1Paper}, we determined a stellar radius of $R_\star\,=\,0.674\,\pm\,0.039$\,$R_\odot$, in agreement with the spectroscopic value derived using \texttt{SpecMatch-emp} (see Table~\ref{Tab:SpecParam}). 

An independent fit of the SED performed with the \texttt{VOSA SED} fitting tool \citep{Bayo2008} yielded a stellar radius of $R_\star\,=\,0.671\,\pm\,0.042$\,$R_\odot$. We also verified our results combining the {\it Gaia} distance, effective temperature, and $V_\mathrm{J}$ magnitude (Table~\ref{tab:parstellar}) with the bolometric correction calculated from the empirical equations by \citep{Torres2010b} and found a radius of $R_\star$\,=\,$0.666\,\pm\,0.065$\,$R_\odot$. Both values are in excellent agreement with our measurement, adding confidence to our results.

We finally converted \teff, $R_\star$, and [Fe/H] into stellar mass $M_\star$ and surface gravity \logg\ using the empirical relations derived by \citet{Torres2010a} coupled to Monte Carlo simulations. \pname\ is a K7\,V star \citep{Pecaut2013} with an effective temperature of \teff\,=\,$4373\,\pm\,57$\,K, a photospheric iron abundance of [Fe/H]\,=\,$+0.03\,\pm\,0.10$\,dex, a mass of $M_\star$\,=\,\smass, and a radius of $R_\star$\,=\,\sradius, yielding a surface gravity of \logg\,=\,$4.584\,\pm\,0.051$~(cgs). The final adopted values are given in Table~\ref{parstable1}. We note that also \logg\ is consistent with our spectroscopic surface gravity (Table~\ref{Tab:SpecParam}).

We used gyrochronology to estimate the age of \pname\ from the relations by \citet{2015MNRAS.450.1787A} and found $740\,\pm\,360$~Myr, suggesting that the star might be relatively young (Table~\ref{parstable1}).

\section{Stellar activity and frequency analysis of the HARPS data}
\label{Sect:Activity}

\pname\ is an active star. As presented in Sect.~\ref{Sect:K2Photometry}, the \ktwo\ light curve of \pname\ displays quasi-periodic modulation with a peak-to-peak amplitude of about 1\% (Fig.~\ref{Fig:K2LC}). The photometric variability is very likely caused by active regions (spots, faculae, and plages) carried across the visible hemisphere of the stellar disk as the star spins about its axis. This is corroborated by the detection of strong emission components in the cores of the Ca\,{\sc ii} H\,\&\,K lines (Fig.~\ref{Fig:CaHK}), from which we derived an average S-index of $0.938\pm0.074$, indicative of a high level of magnetic activity.

The magnetic activity of \pname\ is expected to produce quasi-periodic signals in time-series RV data, commonly referred to as ``stellar jitter''. We used the code \texttt{SOAP2} \citep{2014ApJ...796..132D} to estimate the RV variation induced by stellar activity. From the amplitude of the photometric variability, the spectroscopic parameters, and the rotation period of the star, we calculated an expected RV semi-amplitude variation of $\sim$5-10\,\ms. 

We performed a frequency analysis of our RV measurements to look for the signature of the transiting planet and search for possible activity-induced signals. For this purpose, we did not include the FIES RVs because of the higher uncertainties, the relatively small number of data points, and the need to account for an offset between the FIES and HARPS data-sets.

\begin{figure}
\centering
\includegraphics[width=\linewidth]{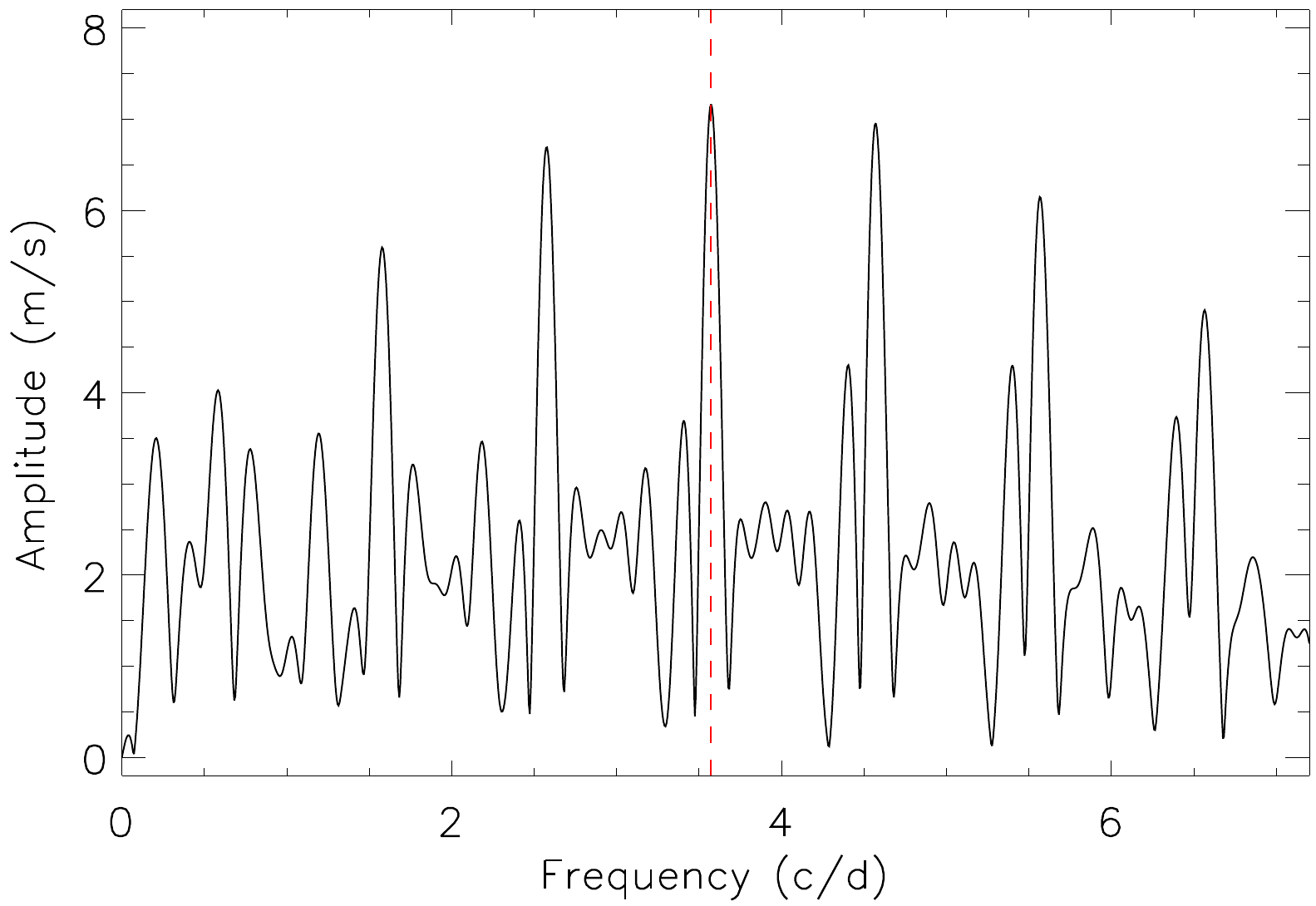}
\caption{Discrete Fourier transform of the HARPS RV measurements. The dashed red line marks the frequency at the orbital period of \pname\,b.} 
\label{Fig:C12_3474_DFT}
\end{figure}

Figure~\ref{Fig:C12_3474_DFT} shows the discrete Fourier transform (DFT) of the 27 HARPS RV measurements calculated using the code \texttt{Period04} \citep{Lenz2004}. The peak with the largest semi-amplitude ($\sim$7\,\ms) is found to be at 3.57 c/d, i.e., the orbital frequency of \pname\,b (0.28 days). We note that the semi-amplitude agrees with the value derived in Sect.~\ref{Sect.:JointAnalysis}. We also note the presence in the DFT of the 1-, 2-, 3-, and 4-day aliases to the left and right of the planetary signal, as expected given the 1-day sampling of our observations.

\begin{figure}
\centering
\includegraphics[width=\linewidth]{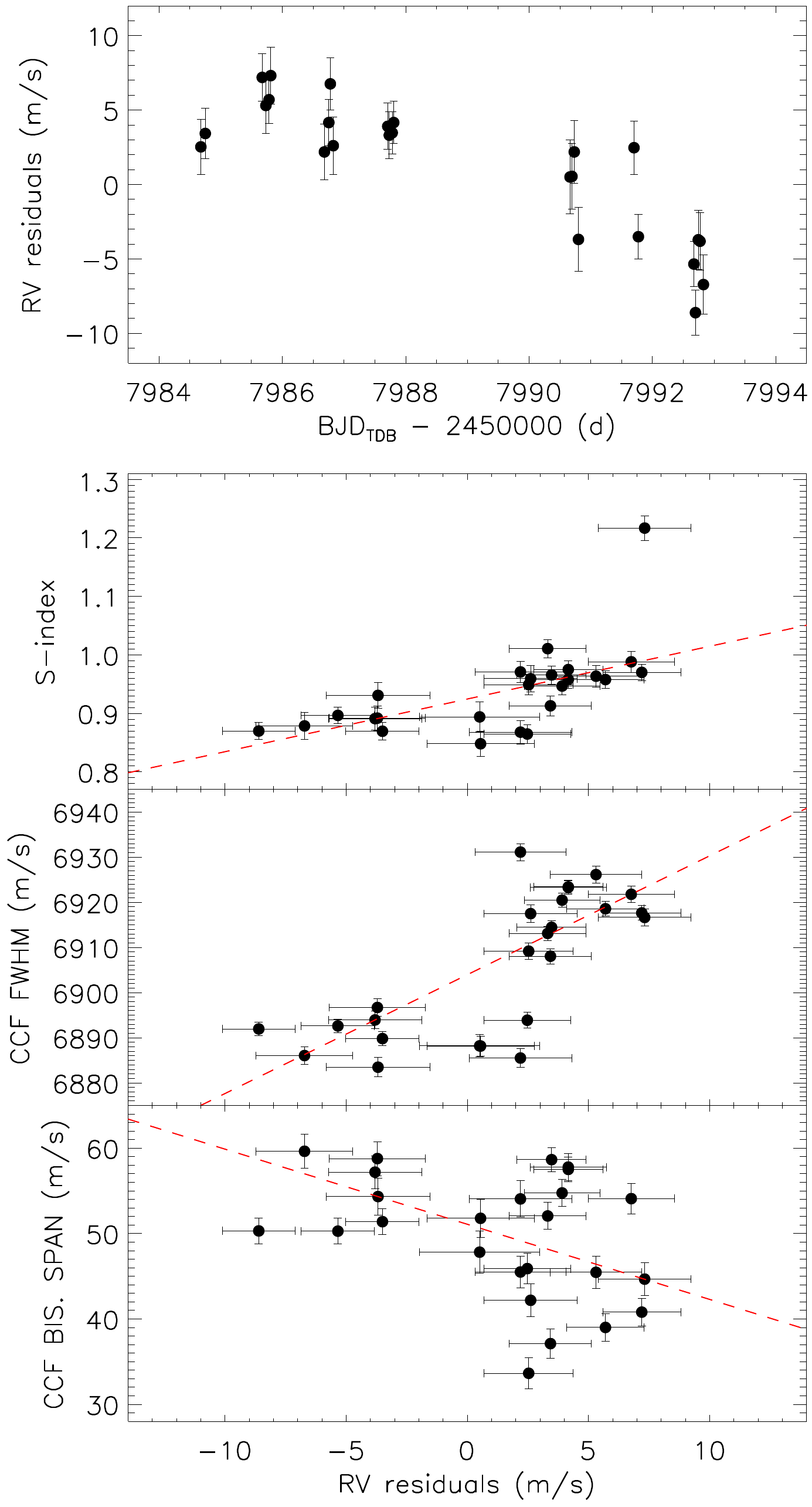}
\caption{\emph{Upper panel}: HARPS RV residuals  following the subtraction of the transiting planet signal versus time. \emph{Lower panel}: Ca\,{\sc ii} H\&K $S$ activity index (top), cross-correlation FWHM (middle), and cross-correlation bisector span (bottom) versus RV residuals. The dashed red lines mark the best fitting linear fit.} 
\label{Fig:activity}
\end{figure}

An additional trend is visible in the RV data after the signal of the transiting planet has been subtracted from the HARPS measurements, as depicted in the upper panel of Fig.~\ref{Fig:activity}. We found a Spearman's rank correlation coefficient of $-0.80$ with a $p$-value of $1.4\times10^{-6}$, strongly suggesting the existence of an additional source of RV variation in our data\footnote{Following \citet{fisher1925}, we adopted a significance level of $p$\,=\,0.05.}. To assess if the source of this additional signal is induced by stellar activity, we looked for possible correlations between the RV residuals and the activity indexes, namely, the Ca\,{\sc ii} H\,\&\,K $S$-index, and the FWHM and bisector span (BIS) of the cross-correlation function (Fig.~\ref{Fig:activity}.). Although the RV residuals and the BIS do not show a significant anti-correlation ($-0.30$ with $p=0.13$), we found significant correlations between the RV residuals and the FWHM (0.73 with $p=3.3\times10^{-5}$), and the RV residuals and the $S$-index (0.72 with $p=4.8\times10^{-5}$). We concluded that the long-term trend observed in the HARPS data is likely caused by the presence of active regions on the photosphere of the star. We will present our approach to filtering out the stellar jitter in the next section.

\section{Data analysis and results}
\label{Sect.:JointAnalysis}

In order to arrive at a robust measurement of the planetary mass despite the additional RV variations induced by stellar activity (see Sect.~\ref{Sect:Activity}), we used three different approaches to fit the data as described below. 

\subsection{Floating chunk offset method}
\label{Sect:FCO_method}

The first method (hereafter M1) is based on the floating chunk offset (FCO)  technique pioneered by \citet{Hatzes2011}, which works well when the orbital period is much shorter than the timescales associated with the signals induced by stellar activity and any additional planets. The FCO method divides a given RV time-series into sub-segments of duration $t$, usually long enough to encompass the RV data collected within a given night. For each sub-segment, an offset is fitted to account for RV signals whose time-scales $\gg t$, assuming that the offset remains constant within one night. From this point of view, \pname\,b is an ideal target to apply the FCO method and remove long-term signals coming from outer companions and stellar activity \citep[see, e.g.,][]{Hatzes2011,Gandolfi2017}. Our observing strategy was tailored to use this technique by acquiring multiple spectra per night (Sect.~\ref{Section:spectroscopy}).

We performed a Markov chain Monte Carlo (MCMC) joint analysis of the transit and RV data using the code \texttt{pyaneti} \citep{pyaneti}. We fitted a Keplerian orbit to the RV data and used the limb-darkened quadratic model by \citet{2002ApJ...580L.171M} for the transit light curves. We integrated the light curve model over 10 steps to account for the \kepler\ long-cadence observation \citep{Kipping2010}. Likelihood and fitted parameters are similar to those described in previous analysis performed with \texttt{pyaneti} \citep[e.g.,][]{Barragan2016,Barragan2017}. We used flat uniform priors for all parameters. Details are given in Table~\ref{parstable1}. We explored the parameter space with 500 Markov Chains to generate a posterior distribution of 250,000 independent points for each parameter. The inferred parameter value and its uncertainty is given by the median and 68.3\% credible interval of the posterior distribution. We did not account for additional jitter terms because $\chi^2/{\rm dof} \approx 1$.

When fitting for an eccentric orbit, the posterior distribution of the eccentricity has a median of 0.06 and a 99\%-confidence upper limit of 0.20. The Bayesian Information Criterion (BIC) favors a circular orbit with a $\Delta\,{\rm BIC} = 8$. Our result is consistent with a circular orbit, as expected for a planet with such a short period. All further analyses were carried out fixing the orbit to be circular. 

We measured a Doppler semi-amplitude of \kbmone[] \ms, which corresponds to a mass of \mpbone.

\subsection{Sinusoidal activity signal modeling}

In the second method (hereafter M2), the RV signal associated with stellar activity is modeled as a coherent sinusoidal signal \citep[e.g.,][]{Pepe2013,Barragan2017}. The \ktwo\ light curve shows the presence of long-livied active regions whose evolution time scale is longer that the rotation period of the star. Since our RV follow-up lasted only $\sim$30\,days, i.e. two stellar rotation periods, we can reasonably assume that the activity-induced RV signal remained coherent within our observing window.

For this method we used \texttt{pyaneti} and performed an MCMC analysis similar to M1. To account for the activity-induced signal at the rotation period of the star, we included an additional sinusoidal signal whose period was constrained with a Gaussian prior centered at $P_\mathrm{rot}$\,=\,14.03\,d with a standard deviation of 0.09\,d (see Sect. \ref{Sect:K2Photometry}). For the  phase and amplitude of the activity signal we adopted uniform priors. 

We first performed a fit including only the planetary signal. This analysis produces a RV $\chi^2/{\rm dof} \approx 5$. When including the extra sinusoidal signal, M2 gives an RV $\chi^2/{\rm dof} \approx 1.3$ and a $\Delta\,{\rm BIC} = 130$ over the previous model. This further proves that RV data cannot be explained by only the planetary signal (cfr. Sect.~\ref{Sect:Activity}). To account for imperfect treatment of the activity-induced variation, we added RV jitter terms to the equation of the likelihood for the FIES and HARPS RV data.

The final inferred Doppler amplitude induced by stellar activity is $5.05_{-0.66}^{+0.72}$\,\ms, which agrees with the prediction made with \texttt{SOAP2} (Sect.~\ref{Sect:Activity}). The RV semi-amplitude variation induced by the planet is \kbmtwo[]\ms, which translates to a planetary mass of \mpbtwo. 

\subsection{Gaussian process}

The third method (hereafter M3) models the correlated noise associated with stellar activity with a Gaussian Process (GP). GP describes stochastic processes with a parametric description of the covariance matrix. GP regression has proven to be successful in modeling the effect of stellar activity for several other exoplanetary systems \citep[see, e.g.,][]{Haywood2014,Grunblatt2015,LM2016}. 

We used the same GP model that was described in detail by \citet{Dai2017}. The list of parameters includes the RV semi-amplitude $K$, the orbital period $P_{\text{orb}}$ and the time of conjunction $t_{\text{c}}$. The model also includes the so-called "hyperparameters" of the quasi-periodic kernel: the covariance amplitude $h$, the correlation timescale $\tau$, the period of the covariance $T$, and $\Gamma$ which specifies the relative contribution between the squared exponential and periodic part of the kernel.



We imposed Gaussian priors on $P_{\text{orb}}$ and $t_{\text{c}}$ using the well-constrained values from {\it K2} transit modeling. We imposed Jeffreys priors on the scale parameters: $h$, $K$, and the jitter parameters. We imposed uniform priors on the systematic offsets $\gamma_{\text{HARPS}}$ and $\gamma_{\text{FIES}}$. Most importantly, we imposed priors on the hyperparameters $\tau$, $\Gamma$ and $T$ based on a Gaussian Process regression of the observed {\it K2} light curve, as described below.

The presence of active regions on the host star coupled with stellar rotation produces quasi-periodic variations in both the measured RV and the flux variation. Given that the activity-induced radial velocity variation and flux variation are generated by similar physical processes, they could be described by Gaussian Processes with similar hyperparameters. Since the photometry has higher precision and better time sampling than the RV data, we used the {\it K2} light curves to constrain the Gaussian Process that describes the observed quasi-periodic variation. We adopted the covariance matrix and the likelihood function described by \citet{Dai2017}. However, since RV and photometric data have different physical dimensions, we replaced $h$ and $\sigma_{\text{jit}}$ with $h_{\text{phot}}$ and $\sigma_{\text{phot}}$, i.e., the amplitude of the quasi-periodic kernel and the white noise component. We imposed a Gaussian prior on $T$ based on the stellar rotation period we measured with the auto-correlation function (Sect.~\ref{Sect:K2Photometry}). Again, Jeffreys priors were imposed on the scale parameters: $h_{\text{phot}}$, $\sigma_{\text{phot}}$, $\tau$ and  $\Gamma$. We imposed a uniform prior on $f_{\text{0}}$, which represents the out-of-transit flux.

We first found the maximum likelihood solution using the Nelder-Mead algorithm implemented in the {\tt Python} package {\tt scipy}. We then sampled the posterior distribution of the various model parameters with the affine-invariant MCMC implemented in the code {\tt emcee} \citep{emcee}. We initialized 100 walkers in the vicinity of the maximum likelihood solution.  We ran the walkers for 5000 links and removed the initial 1000 "burn-in" links. We calculated the Gelman-Rubin potential scale reduction factor, ensuring that it was smaller than 1.03 indicating adequate convergence. We used the median, 16\% and 84\% percentiles of the posterior distribution to summarize the results for the hyperparameters: $\tau$ = $1.9^{+1.3}_{-0.4}$ days,  $\Gamma$ = $2.1^{+1.8}_{-1.0}$. We used these results as Gaussian priors in the subsequent GP analysis of the RV data.

\begin{figure}
\center
\includegraphics[width=0.5\textwidth]{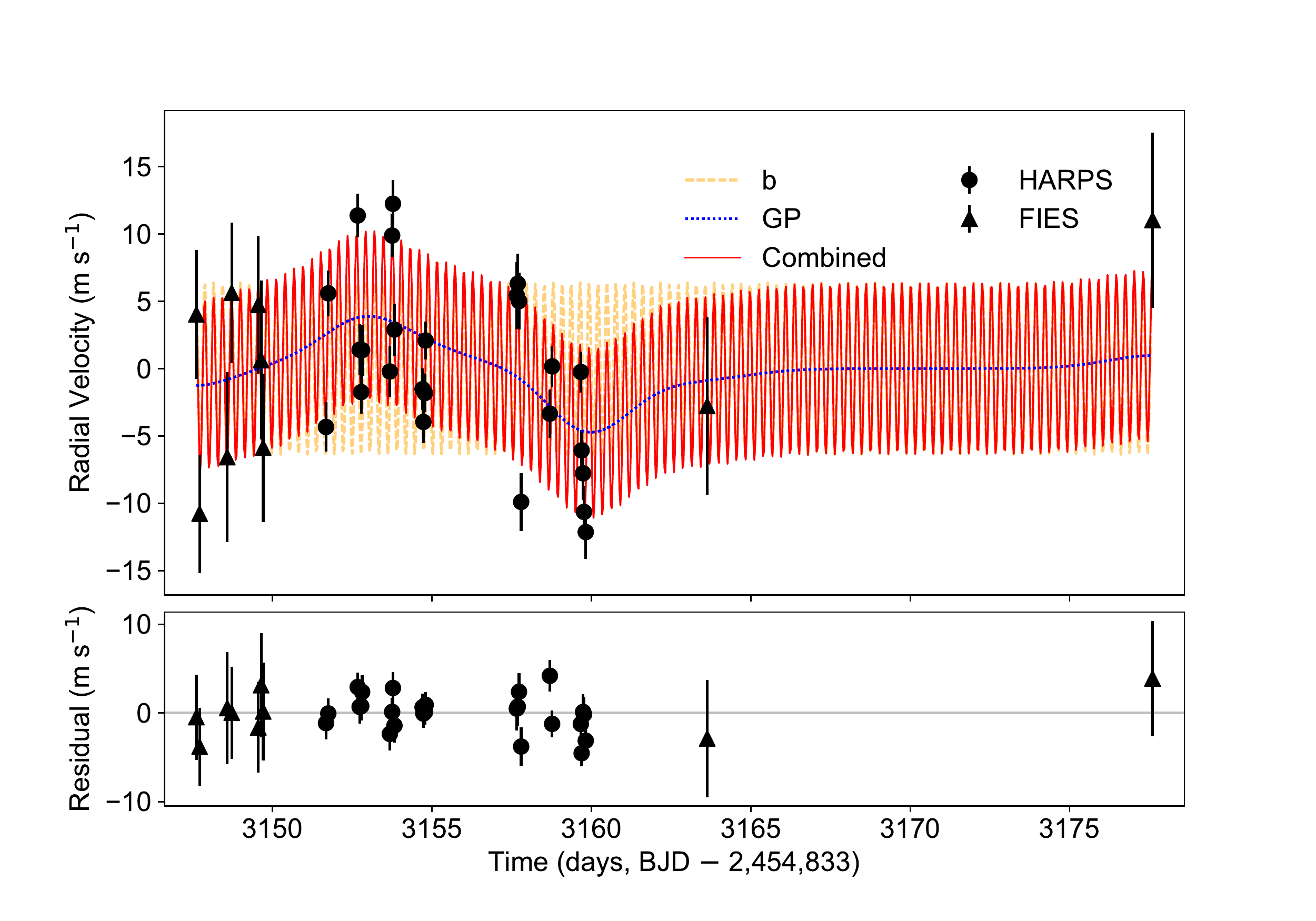} 
\caption{The measured radial velocity variation of EPIC~246393474 from HARPS (circles) and FIES (triangles). The red solid line is the best-fit model including the signal of planet b and the Gaussian Process model of the correlated stellar noise. The yellow dashed line shows the signal of planet b. The blue dotted line shows the Gaussian Process.\label{fig:RV_GP} }
\end{figure}

We analyzed the RV data with Gaussian Process regression by first finding the maximum likelihood solution using the Nelder-Mead algorithm implemented in the {\tt Python} package {\tt scipy}. We sampled the parameter posterior distribution with MCMC using the same procedure as described above. The RV semi-amplitude for planet b was found to be \kbmthree[]~m~s$^{-1}$. This translates into a planetary mass of \mpbthree. Fig.~\ref{fig:RV_GP} shows the measured RV variation of \pname\ and the GP model. We found that the amplitude of the correlated noise is $h_{\text{rv}}$ = $4.0^{+2.2}_{-1.2}$\,m\,s$^{-1}$, which agrees with the value inferred by M2.  

\section{Discussion}

\begin{figure*}
\includegraphics[height=0.30\textwidth]{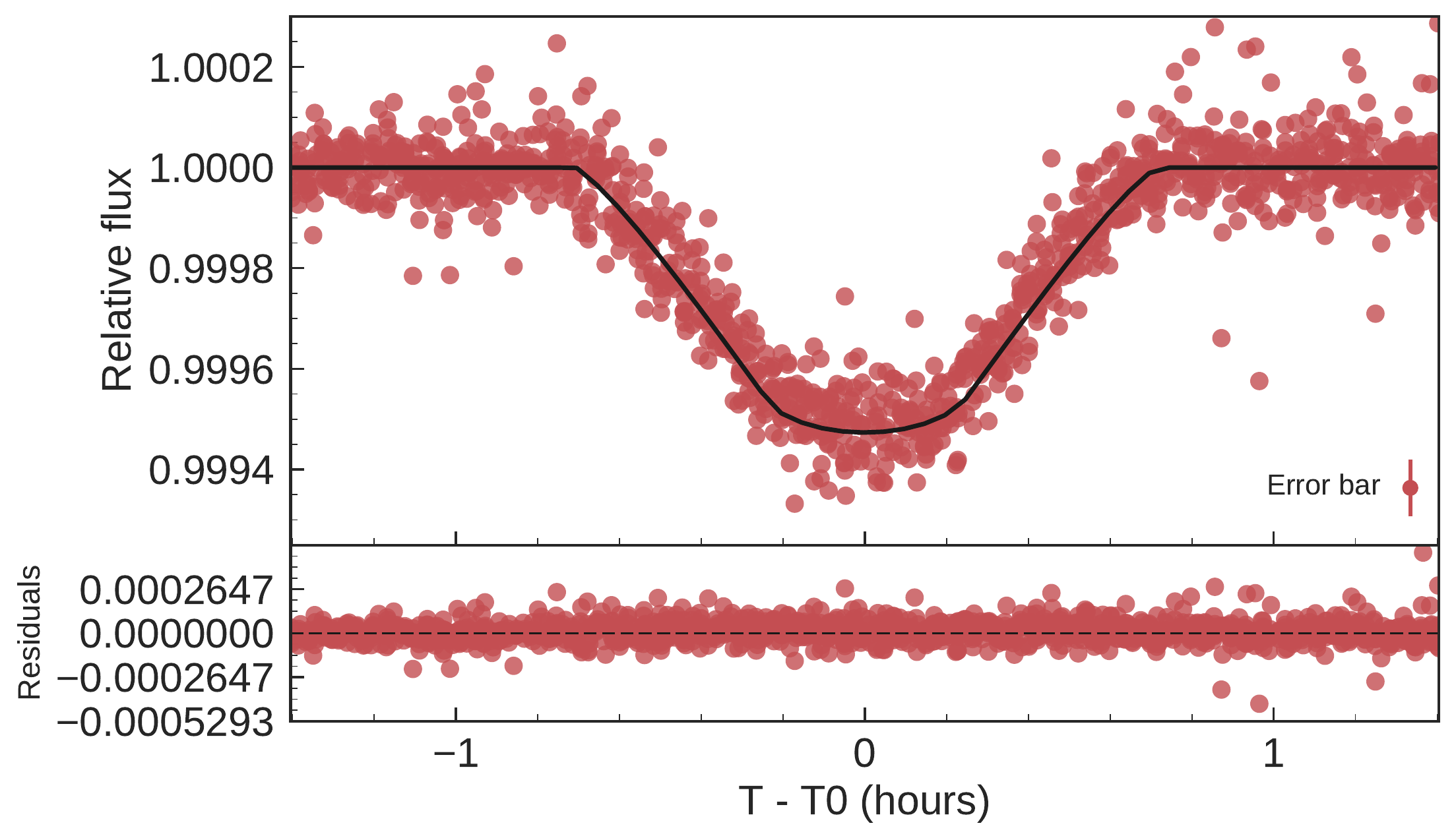} 
\includegraphics[height=0.30\textwidth]{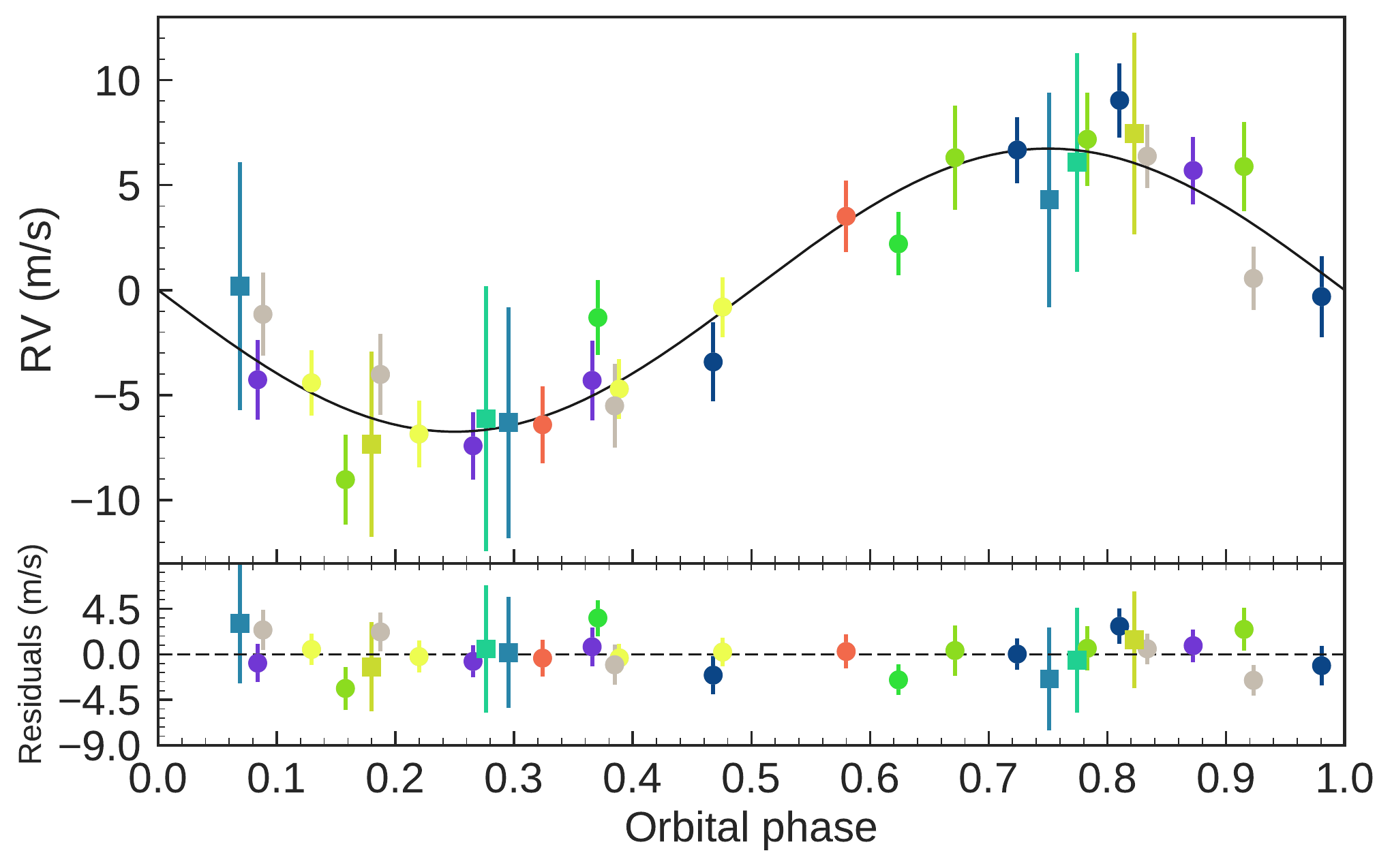} 
\caption{ \emph{Left panel.} Transit light curve folded to the orbital period of \pname\,b and residuals. The red points mark the \ktwo\ data, whereas the thick black line the re-binned best-fitting transit model. \emph{Right panel.} Phase-folded RV curve of \pname\ folded to the orbital period of the planet, as obtained using the FCO method. The best fitting circular solution is marked with a solid black line. HARPS and FIES data are shown with filled circles and squares, respectively. Different colors refer to different nights. The lower panel shows the residuals to the best fitting model.
\label{fig:fits} }
\end{figure*}  

The three techniques used to determine the mass of \pname\,b give results that are consistent to within $\sim$$0.5\sigma$. While we have no reason to prefer one method over the other, we adopted the results of M1 (FCO method), which gives a planetary mass of $M_\mathrm{p}$\,=\,\mpbone\ ($\sim$11$\sigma$ significance). Figure \ref{fig:fits} displays the \ktwo\ and RV measurements, along with the inferred transit and Keplerian models from the FCO method folded to the orbital period of the planet. The parameter estimates are given in Table\,\ref{parstable1}. 

\begin{figure}
\includegraphics[width=0.5\textwidth]{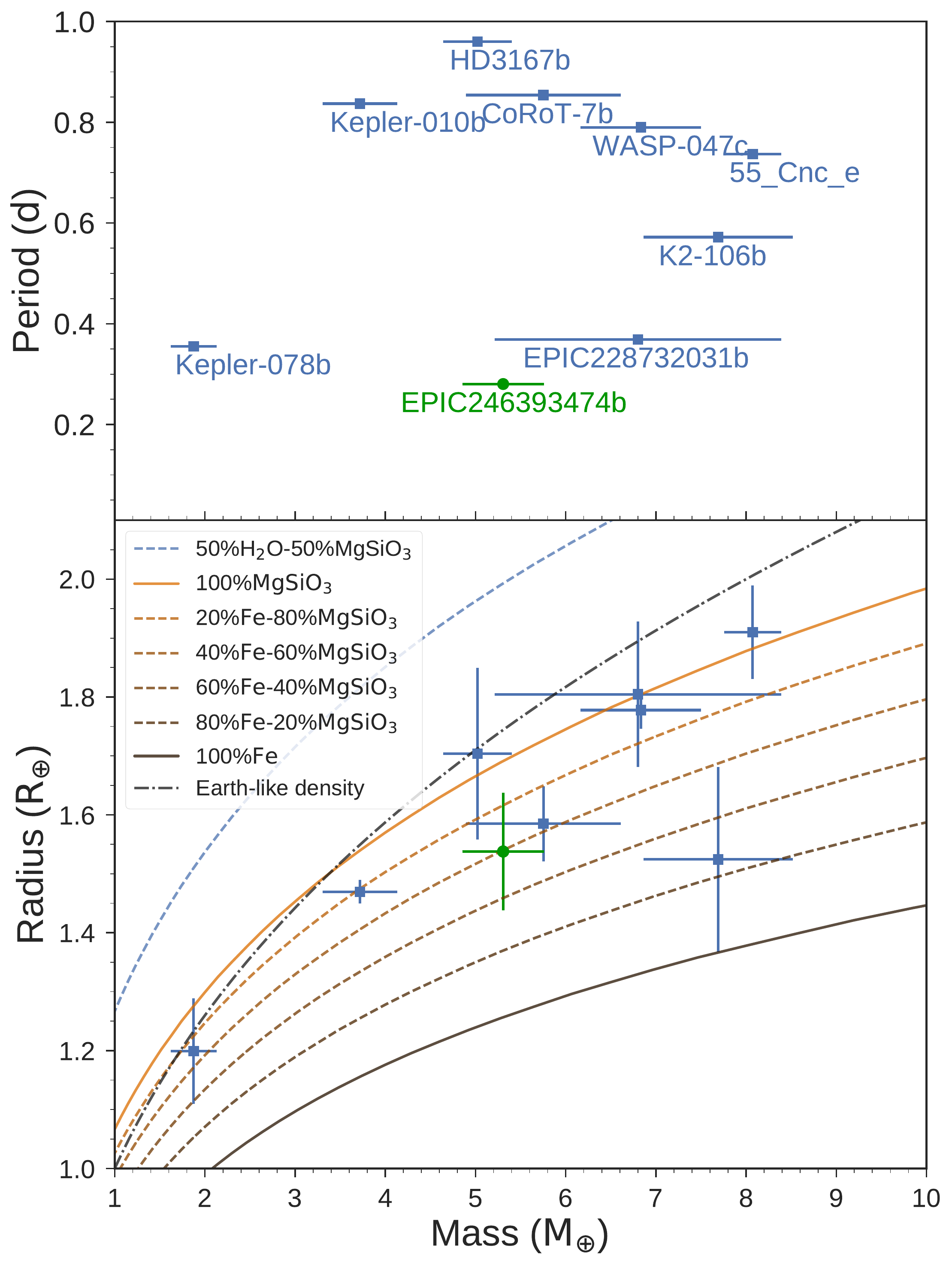}
\caption{{Mass-period (upper panel) and mass-radius (lower panel) diagram for USP planets ($P_{\rm orb}\,<\,1$\,day, $R_\mathrm{p}\,<\,2 R_\oplus$) with measured masses}. The solid green circle marks the position of \pname\,b. USP planets in the literature are marked with blue squares. The composition models from \citet{Zeng2016} are displayed with different lines and colors. The Earth-like density curve is also shown with a dot-dashed black line.
\label{fig:massradius} }
\end{figure}

The upper panel of Fig.~\ref{fig:massradius} shows the mass-period diagram for all the USP planets with directly measured masses. We included 55\,Cnc\,e, CoRoT-7\,b, HD\,3167\,b, K2-106\,b, Kepler-10\,b, Kepler-78\,b and EPIC\,228732031\,b using the planetary masses and radii reported in the TEPCat database\footnote{\url{http://www.astro.keele.ac.uk/jkt/tepcat/}.}. With a period of 0.28\,d (6.7\,h), \pname\,b is the shortest-period planet with a measured mass among all planets known to date. As of October 2017, there are only three transiting exoplanets known to have orbital periods shorter than \pname\,b, namely, Kepler\,70\,b, KOI-1843\,b, and EPIC\,228813918\,b. However, their masses have not yet been measured. The mass of Kepler\,70\,b \citep[$P=0.24\,$d;][]{Charpinet2011} was estimated based on the radius and an assumed mean density. For KOI-1843\,b ($P = 0.18\,$d), the mass was constrained based on the lower limit of the planet's mean density calculated from the requirement that the planet must orbit outside the star's Roche limit \citep{Rappaport2013}. For EPIC\,228813918\,b ($P = 0.18\,$d) \citet{Smith2017b} reported a lower limit  for the planetary mass based on \citet{Rappaport2013}, as well as a $3\sigma$ upper limit based on RV measurements.

The mass of $M_\mathrm{p}$\,=\,\mpbone\ and radius of $R_\mathrm{p}$\,=\,\rpb\ yield a mean density of $\rho_\mathrm{p}$\,=\,\denpb. The lower panel of Fig.~\ref{fig:massradius} shows the mass-radius diagram for USP small transiting planets ($P_{\rm orb}\,<\,1$\,day, $R_\mathrm{p}\,<\,2 R_\oplus$), along with \citet{Zeng2016}'s theoretical models for different compositions.
\citep{Dressing2015} suggested that planets with masses between $1$ and $6~M_{\oplus}$ are consistent with a composition of 17\% Fe and 83\% MgSiO$_3$ (rock). Their sample included 3 USP planets (Kepler-78\,b, Kepler-10\,b and CoRoT-7\,b), two planets with periods 4.3 and 13.8 days, and the Solar System planets Earth and Venus. Figure \ref{fig:massradius} shows a 20\% Fe and 80\% MgSiO$_3$ composition line, similar to the values found by \citet{Dressing2015}. Given its mass and radius, \pname\,b lies close to the 40\%Fe-60\% MgSiO$_3$ compositional model. Within the 1$\sigma$ uncertainties, \pname\,b lies between the 20\%Fe-80\% MgSiO$_3$ and 60\%Fe-40\% MgSiO$_3$ models. If we consider only the 5 USP planets with $M_\mathrm{p}<6~M_\oplus$, we found that no single theoretical curve is consistent with them all. At face value this shows that there is some dispersion in the composition of USP planets.

\begin{figure} 
\includegraphics[width=0.5\textwidth]{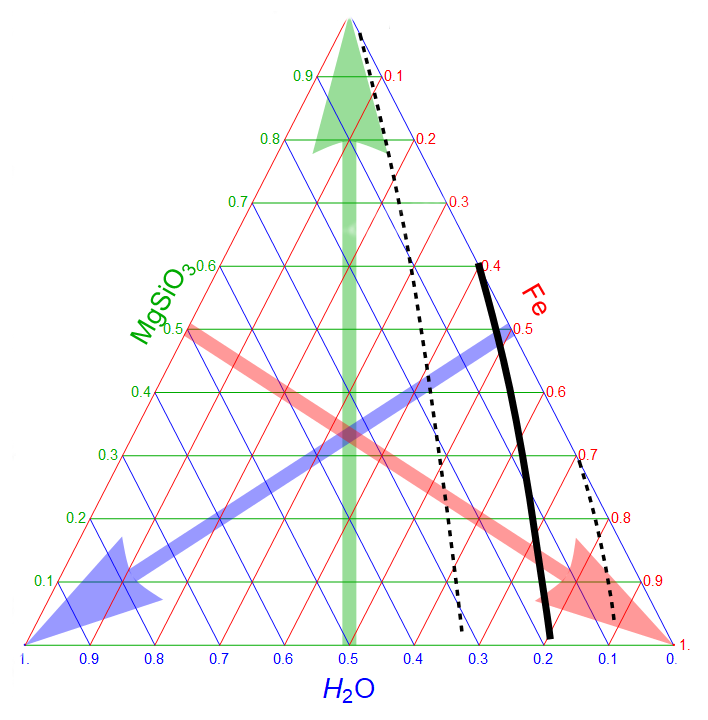}
\caption{{Ternary plot for different planetary compositions}. We show different combinations of water, rock and iron for possible solid planet compositions. The solid and dashed black lines mark the possible position of \pname\,b and the 68\,\% credible intervals. This plot was created using the applet available at \url{https://www.cfa.harvard.edu/~lzeng/manipulateplanet.html}. \label{fig:ternaryplot} }
\end{figure}

We further inferred the composition of \pname\,b using a ternary plot (Figure~\ref{fig:ternaryplot}) for planetary compositions comprising different abundances of H$_2$O, MgSiO$_3$ and Fe \citep{Zeng2013,Zeng2016}. The dashed lines mark the allowed region for \pname\,b, given the 1$\sigma$ uncertainty on the planetary mass and radius.  We first analyzed a water-free model (right-hand side of the triangle). If the planet does not contain H$_2$O, \pname\,b has a composition comprising $\sim$5-70\% iron and $\sim$30-95\% rocks. If the planet does contain H$_2$O, then the maximum water abundance (1$\sigma$ upper limit) cannot exceed $\sim$30\% of the total mass.

Using precise radii for 2024 \kepler\
planets with $P_\mathrm{orb}$$<$100\,d, \citet{Fulton2017} found a deficit of objects with 1.5\,$\lesssim$\,$R_\mathrm{p}$\,$\lesssim$\,2\,$R_\oplus$. This gap divides close-in small planets into two distinct classes: one population comprises planets with $R_\mathrm{p}$$\lesssim$1.5\,$R_\oplus$, the other sub-Neptunes with 2$\lesssim$$R_\mathrm{p}$$\lesssim$\,3.0\,$R_\oplus$. Theoretical models suggest that the observed gap might be due to photo-evaporation \citep[e.g.,][]{Lopez2014,Owen2013}. According to these models, close-in (a\,$\lesssim$\,0.1 AU) planets in the sub-Neptunes regime would lose their atmosphere within a few hundreds Myr due to the intense level of photo-ionizing radiation from their host stars, forming bare rocky cores. \pname\,b receives a stellar radiation of about $~2900\,F_{\oplus}$ (where $F_{\oplus}$ refers to insolation received on Earth), which is more than four times the threshold of $650\,F_\oplus$ needed for planets to undergo photo-evaporation of  H/He envelopes \citep{Lundkvist2016}. This implies that if \pname\,b had an atmosphere, it was lost due to the vicinity to its host star.

\citet{Rappaport2013} pointed out that for planets with orbital periods $\lesssim 6\,{\rm h}$, the mere requirement that the planet is outside the Roche limit leads to an astrophysically relevant lower limit on the planet's mean density. Assuming a constant density, a planet with a period of 6.7\,h would need a minimum density of $\sim$ $3.5\, {\rm g\,cm^{-3}}$ to avoid tidal destruction by the star. This value is smaller than \pname\,b's density lower limit of $5.2\, {\rm g\,cm^{-3}}$. In this case the RV data are more powerful than the Roche-limit consideration, for determining the planet's composition.

\section{Conclusions}

We present the discovery and characterization of \pname\,b, an USP planet transiting an active K7\,V star in \ktwo\ campaign 12. The relatively low stellar mass ($M_\star$\,=\,\smass) and the ultra short orbital period ($P_\mathrm{orb}$\,=\,6.7 hours), along with our multi-visit observing strategy, allowed us to detect the Doppler reflex motion of the star with a significance of about 11$\sigma$. With a mass of $M_\mathrm{p}$\,=\,\mpbone, \pname\,b is the planet with the shortest orbital period and a measured mass known to date.

The planetary density is consistent with a composition made of a mixture of iron and rocks. We estimated that the iron content of \pname\,b cannot exceed $\sim$70\% of the total planetary mass. 

\begin{table*}\footnotesize
\begin{center}
  \caption{Stellar and planetary parameters \label{parstable1}}  
  \begin{tabular}{lcc}
  \hline
  \hline
  \noalign{\smallskip}
  Parameter & Prior$^{(a)}$ & Final value \\
  \noalign{\smallskip}
  \hline
  \noalign{\smallskip}
    \multicolumn{3}{l}{\emph{\bf{Stellar parameters}}} \\
    \noalign{\smallskip}
    Star mass $M_{\star}$ ($M_\odot$) & $\cdots$ &  \smass[] \\
    Star radius $R_{\star}$ ($R_\odot$) & $\cdots$ & \sradius[]  \\
    Stellar density $\rho_\star$ (g\,cm$^{-3}$) & $\cdots$ & $3.05^{+0.61}_{-0.48}$ \\   
    \noalign{\smallskip}
    Stellar density $\rho_\star$ (from light curve, g\,cm$^{-3}$) & $\cdots$ & \densb[] \\
    Effective Temperature $\mathrm{T_{eff}}$ (K) & $\cdots$ & \stemp[] \\
    Surface gravity \logg\ (cgs) & $\cdots$ & $4.584\,\pm\,0.051$ \\ 
    Iron abundance [Fe/H] (dex) & $\cdots$ & $0.03\pm0.10$ \\
    Projected rotational velocity \vsini\ (\kms) & $\cdots$ & $3.0\pm1.7$\\
    Rotational period $P_\mathrm{rot}$ (days) &  $\cdots$ & $14.03\pm0.09$ \\
    Gyrochronological age (Myr) & $\cdots$ & $740\pm360$ \\
    Interstellar extinction $A_\mathrm{V}$ (mag) & $\cdots$ &  $0.01\pm0.02$ \\
    Star distance $d$ (pc) & $\cdots$ & $58.77\pm2.81$ \\
    \noalign{\smallskip}
  \hline
  \noalign{\smallskip}
  \multicolumn{3}{l}{\emph{\bf Model Parameters}} \\
  \noalign{\smallskip}
    Orbital period $P_{\mathrm{orb}}$ (days)  & $\mathcal{U}[0.2802 , 0.2804 ]$ &\Pb[] \\
    Transit epoch $T_0$ (BJD$_\mathrm{TDB}-$2\,450\,000)  & $\mathcal{U}[7738.45 , 7738.47]$ & \Tzerob[]  \\  
    \noalign{\smallskip}
    Scaled semi-major axis $a/R_{\star}$  & $\mathcal{U}[1.1,20]$ & \arb[] \\
    \noalign{\smallskip}
    Scaled planet radius $R_\mathrm{p}/R_{\star}$  &$\mathcal{U}[0.0,0.1]$ & \rrb[]  \\
    \noalign{\smallskip}
    Impact parameter, $b$ & $\mathcal{U}[-1,1]$ & \bb[] \\
    \noalign{\smallskip}
	$\sqrt{e} \sin \omega_\star$  & $\mathcal{F}[0]$ & 0  \\
    $\sqrt{e} \cos \omega_\star$ &  $\mathcal{F}[0]$ &0   \\
    Parameterized limb-darkening coefficient $q_1^{(b)}$  & $\mathcal{U}[0,1]$ & \qone  \\
    \noalign{\smallskip}
    Parameterized limb-darkening coefficient $q_2^{(b)}$  &$\mathcal{U}[0,1]$ & \qtwo \\  
    Doppler semi-amplitude variation $K$ (m s$^{-1}$) & $\mathcal{U}[0,50]$ & \kbmone[] \\
    \noalign{\smallskip}
    \hline
    \noalign{\smallskip}
    \multicolumn{3}{l}{\emph{\bf Derived Parameters}} \\
    \noalign{\smallskip}
    Planet mass $M_\mathrm{p}$ ($M_{\rm \oplus}$) & $\cdots$ & \mpbone[]  \\
    Planet radius $R_\mathrm{p}$ ($R_{\rm \oplus}$) & $\cdots$ & \rpb[] \\
    \noalign{\smallskip}
    Planet density $\rho_{\rm p}$ (g\,cm$^{-3}$) & $\cdots$ & \denpb[] \\
    \noalign{\smallskip}
	Semi-major axis of the planetary orbit $a$ (AU) &  $\cdots$ & \ab[]  \\
    Eccentricity$^{(\mathrm{c})}$ $e$                                & $\cdots$ & 0 \\
    Orbit inclination along the line-of-sight $i_\mathrm{p}$ ($^{\circ}$) & $\cdots$ & \ib[] \\
    Transit duration $\tau_{14}$ (hours) & $\cdots$ & \ttotb[] \\
    Equilibrium temperature$^{(\mathrm{d})}$  $T_\mathrm{eq}$ (K)  &  $\cdots$ & \Tequib[] \\
    \noalign{\smallskip}
    Linear limb-darkening coefficient $u_1$ & $\cdots$ & \uone \\
    \noalign{\smallskip}
    Quadratic limb-darkening coefficient $u_2$ & $\cdots$ & \utwo\\
   \noalign{\smallskip}
  \hline
  \end{tabular}
  \begin{tablenotes}\footnotesize
  \item \emph{Note} -- $^{(\mathrm{a})}$ $\mathcal{U}[a,b]$ refers to uniform priors between $a$ and $b$, $\mathcal{F}[a]$ to a fixed $a$ value. $^{(\mathrm{b})}$ $q_1$ and $q_2$ as defined by \citet{Kipping2013}. $^{(c)}$~Fixed (see Sect.~\ref{Sect:FCO_method}). $^{(d)}$~Assuming albedo = 0. 
\end{tablenotes}
\end{center}
\end{table*}

\begin{acknowledgements}
We are very grateful to the NOT and ESO staff members for their unique and superb support during the observations. Data presented herein were obtained at the WIYN Observatory from telescope time allocated to NN-EXPLORE through the scientific partnership of the National Aeronautics and Space Administration, the National Science Foundation, and the National Optical Astronomy Observatory, obtained as part of an approved NOAO observing program (P.I. Livingston, proposal ID 2017B-0334). NESSI was built at the Ames Research Center by Steve B. Howell, Nic Scott, Elliott P. Horch, and Emmett Quigley. D.\,G. gratefully acknowledges the financial support of the \emph{Programma Giovani Ricercatori -- Rita Levi Montalcini -- Rientro dei Cervelli (2012)} awarded by the Italian Ministry of Education, Universities and Research (MIUR). This work is partly financed by the Spanish Ministry of Economics and Competitiveness through projects ESP2014-57495-C2-1-R and ESP2016-80435-C2-2-R. M.\,F. and C.\,M.\,P. acknowledge generous support from the Swedish National Space Board. Sz.\,Cs. thanks the Hungarian OTKA Grant K113117.

This paper includes data collected by the \kepler\ mission. Funding for the \kepler\ mission is provided by the NASA Science Mission directorate. This work has made use of data from the European Space Agency (ESA) mission {\it Gaia} (\url{https://www.cosmos.esa.int/gaia}), processed by the {\it Gaia} Data Processing and Analysis Consortium (DPAC, \url{https://www.cosmos.esa.int/web/gaia/dpac/consortium}). Funding for the DPAC has been provided by national institutions, in particular the institutions participating in the {\it Gaia} Multilateral Agreement. This publication makes use of VOSA, developed under the Spanish Virtual Observatory project supported from the Spanish MICINN through grant AyA2011-24052.
\end{acknowledgements}


\begin{table*}[!t]
\begin{center}
\caption{FIES and HARPS measurements of \pname.}
\begin{tabular}{lcccccccc}
\hline
\hline
\noalign{\smallskip}
BJD$_\mathrm{TDB}$ &   RV   & $\sigma_\mathrm{RV}$  & BIS        & FWHM   & S-index & $\sigma_\mathrm{S-index}$ & $T_\mathrm{exp}$ & S/N         \\
 -2\,450\,000      & (\kms) &        (\kms)         &    (\kms)  & (\kms) &   &                     & (s)                & @5500~\AA   \\
\noalign{\smallskip}
\hline
\noalign{\smallskip}
\multicolumn{4}{l}{FIES} \\
\noalign{\smallskip}
7980.614462  &  ~~0.0000  &  0.0048 & \dotfill & \dotfill & \dotfill & \dotfill & 2700 & 44.1 \\   
7980.714496  &  $-$0.0148 &  0.0044 & \dotfill & \dotfill & \dotfill & \dotfill & 2700 & 43.2 \\     
7981.582595  &  $-$0.0106 &  0.0063 & \dotfill & \dotfill & \dotfill & \dotfill & 2700 & 38.4 \\     
7981.722148  &  ~~0.0016  &  0.0052 & \dotfill & \dotfill & \dotfill & \dotfill & 2700 & 41.0 \\     
7982.556610  &  ~~0.0007  &  0.0051 & \dotfill & \dotfill & \dotfill & \dotfill & 2700 & 40.9 \\     
7982.645724  &  $-$0.0034 &  0.0059 & \dotfill & \dotfill & \dotfill & \dotfill & 2700 & 40.0 \\     
7982.709129  &  $-$0.0099 &  0.0055 & \dotfill & \dotfill & \dotfill & \dotfill & 2700 & 44.0 \\     
7996.636413  &  $-$0.0068 &  0.0066 & \dotfill & \dotfill & \dotfill & \dotfill & 2700 & 38.1 \\     
8010.600636  &  ~~0.0070  &  0.0065 & \dotfill & \dotfill & \dotfill & \dotfill & 2700 & 38.2 \\  
\noalign{\smallskip}
\hline
\noalign{\smallskip}
\multicolumn{4}{l}{HARPS} \\
\noalign{\smallskip}
 7984.679455 & $-$3.3841 & 0.0018 &  0.0336 & 6.9092 &  0.949 & 0.018 & 2100 & 53.1 \\
 7984.751182 & $-$3.3742 & 0.0017 &  0.0371 & 6.9081 &  0.913 & 0.017 & 2100 & 57.3 \\
 7985.674124 & $-$3.3684 & 0.0016 &  0.0408 & 6.9177 &  0.970 & 0.014 & 3600 & 59.5 \\
 7985.733443 & $-$3.3784 & 0.0019 &  0.0455 & 6.9261 &  0.964 & 0.019 & 2100 & 52.1 \\
 7985.784325 & $-$3.3815 & 0.0016 &  0.0390 & 6.9186 &  0.958 & 0.015 & 2700 & 60.5 \\
 7985.812451 & $-$3.3784 & 0.0019 &  0.0447 & 6.9167 &  1.217 & 0.021 & 2400 & 52.6 \\
 7986.682021 & $-$3.3800 & 0.0019 &  0.0455 & 6.9311 &  0.971 & 0.018 & 2100 & 52.5 \\
 7986.753853 & $-$3.3699 & 0.0016 &  0.0578 & 6.9233 &  0.958 & 0.015 & 2100 & 61.7 \\
 7986.778032 & $-$3.3675 & 0.0018 &  0.0541 & 6.9218 &  0.988 & 0.018 & 2100 & 55.6 \\
 7986.825742 & $-$3.3769 & 0.0019 &  0.0422 & 6.9175 &  0.959 & 0.022 & 2100 & 52.1 \\
 7987.708413 & $-$3.3813 & 0.0016 &  0.0548 & 6.9205 &  0.947 & 0.015 & 2100 & 62.4 \\
 7987.733807 & $-$3.3837 & 0.0016 &  0.0521 & 6.9131 &  1.011 & 0.016 & 2100 & 61.4 \\
 7987.781124 & $-$3.3816 & 0.0014 &  0.0587 & 6.9145 &  0.966 & 0.016 & 2100 & 67.8 \\
 7987.805535 & $-$3.3777 & 0.0014 &  0.0575 & 6.9234 &  0.975 & 0.015 & 2100 & 68.4 \\
 7990.663651 & $-$3.3743 & 0.0025 &  0.0478 & 6.8882 &  0.894 & 0.026 & 2400 & 41.3 \\
 7990.694913 & $-$3.3735 & 0.0022 &  0.0518 & 6.8882 &  0.848 & 0.022 & 3000 & 45.1 \\
 7990.731940 & $-$3.3748 & 0.0021 &  0.0541 & 6.8855 &  0.868 & 0.020 & 3000 & 46.6 \\
 7990.799975 & $-$3.3897 & 0.0021 &  0.0543 & 6.8835 &  0.931 & 0.022 & 2400 & 46.2 \\
 7991.700580 & $-$3.3831 & 0.0018 &  0.0459 & 6.8939 &  0.865 & 0.016 & 3600 & 53.7 \\
 7991.771590 & $-$3.3796 & 0.0015 &  0.0514 & 6.8898 &  0.870 & 0.015 & 3000 & 63.6 \\
 7992.671326 & $-$3.3800 & 0.0015 &  0.0503 & 6.8926 &  0.897 & 0.014 & 2100 & 63.0 \\
 7992.696419 & $-$3.3858 & 0.0015 &  0.0503 & 6.8920 &  0.870 & 0.014 & 2100 & 63.9 \\
 7992.742740 & $-$3.3875 & 0.0020 &  0.0588 & 6.8967 &  0.892 & 0.021 & 2100 & 50.0 \\
 7992.770507 & $-$3.3904 & 0.0019 &  0.0572 & 6.8939 &  0.891 & 0.020 & 2400 & 51.5 \\
 7992.825821 & $-$3.3919 & 0.0020 &  0.0596 & 6.8861 &  0.879 & 0.023 & 2100 & 50.4 \\
\noalign{\smallskip}
\hline
\end{tabular}
\label{RV}
\end{center}
\end{table*}

\bibliographystyle{aa} 
\bibliography{bibs} 

\end{document}